\documentclass[journal=jacsat,manuscript=article]{achemso}

\usepackage[version=3]{mhchem}
\usepackage[colorlinks=true, allcolors=blue]{hyperref}
\usepackage{soul}
\usepackage[T1]{fontenc}
\usepackage[utf8]{inputenc}
\usepackage{xcolor}
\author{Horacio Serna}
\affiliation[IChF]
{Institute of Physical Chemistry Polish Academy of Sciences, Kasprzaka 44/52 01-224 Warsaw, Poland.}

\author{Wojciech T. G\'o\'zd\'z}
\email{wtg@ichf.edu.pl}
\affiliation[IChF]
{Institute of Physical Chemistry Polish Academy of Sciences, Kasprzaka 44/52 01-224 Warsaw, Poland.}

\title[An \textsf{achemso} demo]
  {The influence of obstacles on collective motion of self-propelled objects. }
  
\abbreviations{}

\begin{document}


\begin{abstract}
We investigate the influence of obstacles on the collective motion of self propelled objects in the framework of the Vicsek model. The obstacles are arranged in a square lattice and have circular shape. We show that by increasing the radius of the obstacles the collective motion of the self propelled object can be altered from super  diffusive to diffusive. For obstacles with small radius, the system is composed of large clusters moving in one direction, for larger radius, the system is composed of small clusters moving randomly in different directions.

\end{abstract}

\section{Introduction}\label{intro}

Collective motion of self propelled objects\cite{gompper20202020} is a fasctnating phenomenon . It takes palce at various time and length scales.  For example colonies of bacteria, swarms of bees, school of fish and flocks of birds can move collectively.
Vicsek model is one of the simplest mathematical models to study the collective motion of self propelled particles \cite{vicsek1995novel}.  
Polar liquid, band and disordered phases were discovered in Vicsek model for different values of the noise amplitude \cite{chate2008collective}. Recently, a novel cross sea phase\cite{kursten2020dry} was reported for noise amplitudes between the polar liquid and band phases.

In recent years, many works have been devoted to Vicsek model in either the presence of obstacles or confined into different geometries, revealing interesting dynamical behaviour. For instance, it has been determined that in a continuous version of the Vicsek model, the confinement into one-dimensional channels, squares and circles can lead to spatial coherence of swarms \cite{armbruster2017swarming}. On the other hand, an on-lattice Vicsek model was studied in the same confining geometries, showing alignment of self propelled  particles parallel to the boundaries at low noise amplitudes for channels and circles and trapping at the corners for squares \cite{kuhn2021lattice}.
The influence of randomly distributed obstacles on the collective motion of particles with aligning interactions has been studied with simulation and theoretical approaches \cite{chepizhko2013diffusion,chepizhko2015active}. 
The effects of fixed and mobile obstacles on the behaviour of  Vicsek disks was considered in a recent work \cite{martinez2018collective}. 
It was also shown that the presence of a small obstacle at the centre of a large system of point-like Vicsek particles can reverse the direction of movement of a polar flock \cite{codina2022small}.
Trapping of self propelled point-like Vicsek particles by arrays of asymmetric V-shaped obstacles was also recently investigated  \cite{martinez2020trapping}.
The effects of topology and geometry on the collective motion in Vicsek model was studied by confining the system into different assemblies of circular obstacles \cite{mcclure2022effect}. Interestingly, the authors found that the presence of obstacles not only hinders the alignment of particles within the system but induces sub-diffusive behaviour.

In this article, we study the collective motion in a square lattice of circular obstacles. We analyze the effects of both the packing fraction of obstacles and the separation between them (which is equivalent to varying the number density of obstacles). In contrast to previous works we keep the noise amplitude fixed and focus on the influence of different arrangement and size of obstacles on the collective motion of self propelled objects.

\section{Model and Methods}\label{model}
The collective motion of self of self-propelled particles is studied within the framework of the Vicsek model~\cite{vicsek1995novel}. The movement of the particles is governed by the following equation
\begin{equation}
 \boldsymbol{r}_i(t + \delta t) = \boldsymbol{r}_i (t) + \boldsymbol{v}_i\delta t , 
 \label{e:VicsekX}
\end{equation}
where $\boldsymbol{r}_i(t)$ and $\boldsymbol{v}_i$ denote the position and velocity of a particle $i$ at time $t$. The time step is set to $\delta t  = 1.0$. The speed, $v_0$, of each particle is constant. The angle formed by the velocity with the $x$ axis is denoted by $\theta_i(t)$. The angle $\theta_i(t + \delta t)$ at the next time step is calculated according to the following rule 
\begin{equation}
 \theta_i(t + \delta t) = \langle \theta (t) \rangle _{r_c} + \Delta\theta 
 \label{e:VicsekTheta}
 \end{equation}
where $\langle \theta(t) \rangle_{r_c}$ is an average angle between all the velocity vectors with the horizontal axis,   contained inside a circle of radius $r_c$  with the center set on the particle $i$ given by its coordinates $\boldsymbol{r}_i$

The term $\Delta \theta$ is a random angle uniformly distributed in the interval $[-\zeta/2,\zeta/2]$, where $\zeta$ is the noise amplitude. We consider the particles movement at low speed limit, low noise amplitudes, and low number density. Thus, we set the model parameters to $v_0 = 0.30$, $\zeta = 1.0 rad$ $(\approx 57.3^0)$, and $\rho = N/L^2 = 0.50$, being $N$ the number of particles and $L$ the side length of the simulation box.  For this set of parameters, the system exhibits collective motion in the form of big swarming clusters in bulk (See Supplementary Movie M1).

\begin{figure}[H]
\centering
\includegraphics[width=1\textwidth]{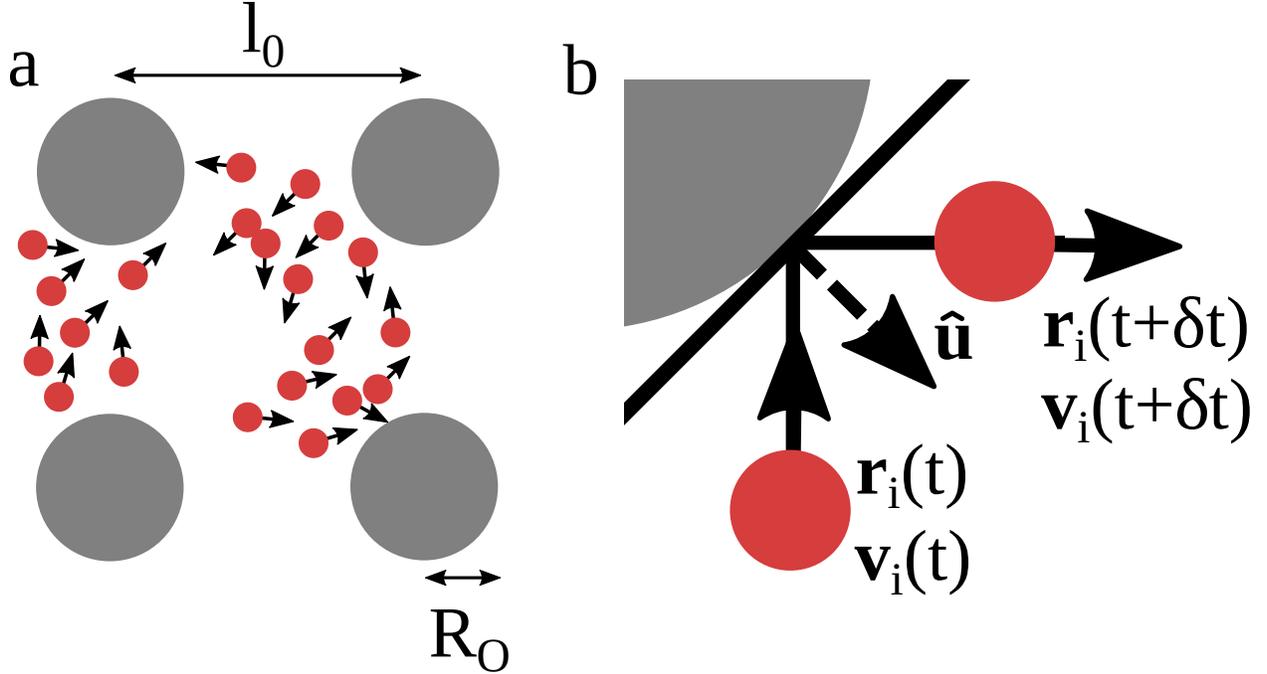}
\caption{\label{fig:Scheme}\textbf{(a)} Scheme of the system. The gray and red circles represent the obstacles and particles, respectively. $l_0$ is the lattice constant of the square lattice and $R_O$ is the radius of the obstacles. \textbf{(b)} Representation of the elastic collisions of a particle with an obstacle within a time step.}   
\end{figure} 

We consider a square lattice of fixed circular obstacles ( See Figure \ref{fig:Scheme}a). The simulations are performed in a periodic square box with $L=300r_c$. The collisions of particles with obstacles are elastic. The  velocity after the collision is calculated according to the following formula $\boldsymbol{v}_i = v_0(\boldsymbol{\hat{v}}_i - 2 (\boldsymbol{\hat{v}}_i \cdot \boldsymbol{\hat{u}})\boldsymbol{\hat{u}})$, where $\boldsymbol{\hat{v}}_i$ is the unit vector parallel to the velocity vector of particle $i$ before the collision and $\boldsymbol{\hat{u}}$ is the unit vector normal to the obstacle's surface at the contact point. To calculate the contact point, we calculate the fraction of time step that the particle requires to arrive to the obstacle's boundary. In the remaining time step the particle travels with the velocity $\boldsymbol{v}_i(t+ \delta t)$ after the collision (see Figure \ref{fig:Scheme}b). We investigate arrays with $N_O = 49$, $225$, $484$ and $900$ obstacles, corresponding to lattice constants, $l_0 = \sqrt{N_O}/L = 42.86r_c$, $20.0r_c$,$13.64r_c$ and $10.0r_c$, respectively. We study the influence of the packing fraction of obstacles by gradually increasing their radius, $R_O$, for each array.

The structure of the system is monitored during the simulations by calculating the order parameter which measures the degree of alignment of the velocities of all particles~\cite{gregoire2004onset}:
\begin{equation}
 \nu_a= \frac{1}{N}\sqrt{\left( \sum_{j=1}^{N} \cos{\theta_j} \right)^2 + \left(\sum_{j=1}^{N} \sin{\theta_j}\right)^2}. 
 \label{e:OP}
 \end{equation}
Average value of the order parameter calculated during the simulation $\langle \nu_a \rangle$, is used to identify structural transition induced by the change of the packing fraction of obstacles, $\eta = \pi N_OR^2_O/L^2$, in a similar way  as it was done by increasing the noise amplitude in the bulk system in the original Vicsek model\cite{vicsek1995novel} and its variations\cite{chate2008collective,chate2008modeling,gregoire2004onset}.  
The simulations are run for $5\times10^5$ time steps, $10^5$ for the system to reach the steady state and the rest for averaging properties. 

To characterize the dynamic behaviour of the system, we compute the mean square displacement,
\begin{equation}
 \left\langle\Delta r(t)^2\right\rangle = \left\langle\frac{1}{N}\sum_{i=1}^{N}\left(\boldsymbol{r}_i(t) - \boldsymbol{r}_i(0)\right)^2\right\rangle
 \label{e:MSD},
 \end{equation}
and the average number of collisions with obstacles per time step, $\langle C\rangle$. The averages are taken at each time step during $5\times10^5$ time steps at the steady state for all the particles.

\section{Results and discussion}\label{results}

\begin{figure}[H]
\centering
\includegraphics[width=1\textwidth]{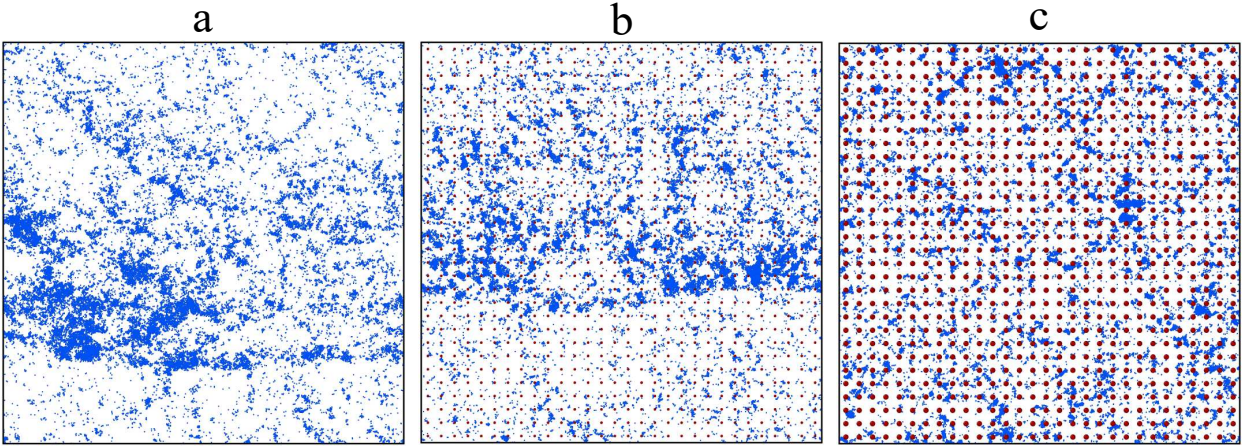}
\caption{\label{fig:Density} The snapshots from the simulations representing different states of the system. The parameters of the model are $v_0 = 0.30$, $\zeta = 1.0rad$, $\rho = 0.50$,  and the system size $L=300r_c$.  
(a) Large clusters in bulk. Systems with obstacles are composed of 30 obstacles per edge ($l_0 = 10.0r_c$), (b) large clusters passing through small obstacles ($R_O = 1.0r_c$, $\eta = 0.0314$ ), (c) small clusters  moving between relatively large obstacles ($R_O = 2.0r_c$, $\eta = 0.1257$). 
}   
\end{figure}

For the chosen parameters of the noise  $\xi=1$, the speed $v_0=0.3$, and the density $\rho=0.5$, the particles form large clusters moving collectively in a preferred direction in bulk. 
For confined systems, we can identify two structural regimes for all four arrays of obstacles. In the first regime where the obstacles are relatively small, we observe large clusters moving in one direction along y or x axis. For larger obstacle smaller clusters are formed, moving in different directions.   
These theree states are illustrated in Figure \ref{fig:Density}.

For small values of the packing fraction, $\eta$, ( small obstacles ) the structure of the systems is similar to the bulk structure. It is interesting to note that even very small obstacles enforce the movement of clusters along the x or y axis. Such behavior results form passing the particles through a square lattice of obstacles.  In this state the system is composed of a few big clusters moving in the same direction and very small clusters and isolated particles. The order parameter, $\nu_a$,  has relative large values for such structure.    

For larger values of the packing fraction, $\eta$, (bigger obstacles) the clusters become smaller and move in different directions. The order parameter, $\nu_a$,  has relative small values for such structures. 
Short movies of the cases presented in Figure \ref{fig:Density} are available in the Supplementary Information( Movie 1 (a). Movie 2 (b) and Movie 3 (c)).
\begin{figure}[H]
\centering
\includegraphics[width=0.5\textwidth]{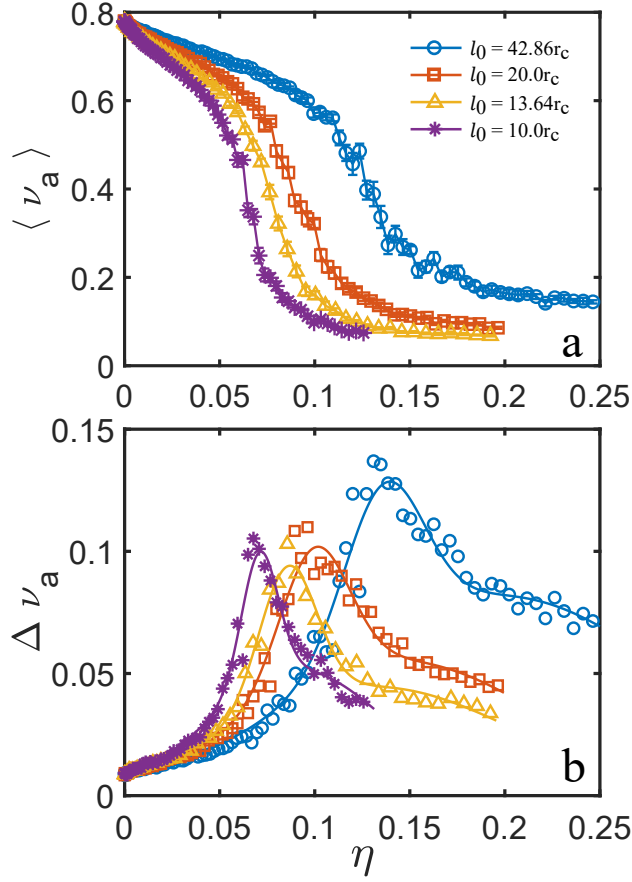}
\caption{\label{fig:OP} \textbf{(a)}The average order parameter, $\langle \nu_a\rangle$, as a function of the packing fraction of obstacles, $\eta$, for all the arrays considered. Solid lines connect the data as a guide for the eye. \textbf{(b)} The fluctuations of order parameter, $\nu_a$, as a function of the packing fraction of the obstacles, $\eta$, for all the arrays considered. Solid lines represent fit curves to double Gaussian functions.}  
\end{figure}

To investigate the alignment of the velocity of the particles, we calculate the average order parameter, $\langle \nu_a\rangle$, as a function of the packing fraction, $\eta$, according to Eqn.~\ref{e:OP}. Additionally, we compute the fluctuations of the order parameter, $\Delta \nu_a  = \langle \nu_a^2 \rangle - \langle \nu_a\rangle^2 $. The results are presented in Figure~\ref{fig:OP}. We observe that for all studied systems, upon increasing the packing fraction of obstacles, the average order parameter, $\langle \nu_a\rangle$, decreases (See Figure \ref{fig:OP}a). The decrease of the order parameter indicates a structural transition between two distinct regimes characterized by big clusters moving in one direction and small clusters moving in different directions.  The slope of the curve suddenly change for certain values of $\eta$, indicating the transition from the first regime to the second regime. The value of the packing fraction at the inflection point can be used to identify the critical packing fraction for this transition. The larger the density of the obstacles the smaller the value of the critical packing fraction (See Figure \ref{fig:OP}a). The region of the transition  is also characterized by increased values of the fluctuations of the order parameter, $\Delta \nu_a$, as shown in Figure \ref{fig:OP}b.  $\Delta \nu_a$ shows a clear peak that signals a the transition. The data for studied systems can be fitted with a double Gaussian function as shown in Figure \ref{fig:OP}b. We take the position of the peaks as a reference to define the critical packing fraction, $\eta_t$. The values of $\eta_t$ and the corresponding $\langle \nu_a \rangle$ are presented in Table \ref{tab:transition}.
\begin{table}
\begin{tabular}{||c | c | c | c||} 
 \hline
 System & $\eta_t$ & $R_t$ & $\langle \nu_a\rangle_t$\\ [0.5ex] 
 \hline\hline
 $l_0 = 42.86r_c$ & $\approx 0.1395$ & $\approx 9.0310 $ & $\approx 0.2735$\\ 
 \hline
$l_0 = 20.00r_c$ & $\approx 0.1018$ & $\approx 3.6002$& $\approx 0.2505$\\
 \hline
 $l_0 = 13.64r_c$ & $\approx 0.0870$ & $\approx 2.2692$ & $\approx 0.2642$\\
 \hline
 $l_0 = 10.00r_c$ & $\approx 0.0718$ & $\approx 1.5117$ & $\approx 0.2496$\\ [1ex] 
 \hline
\end{tabular}
\caption{\label{tab:transition} The critical packing fraction, $\eta_t$, estimated for the studied systems. $l_0$ is the lattice constant of the array of obstacles, $R_t$ is the critical radius of obstacles and $\langle \nu_a \rangle_t$ is the critical average order parameter   }
\end{table}
These results show a new type of the  transition in Vicsek-like models, induced by increasing the packing fraction of obstacles arranged in a square lattice. 

\begin{figure}[htpb]
\centering
\includegraphics[width=0.4\textwidth]{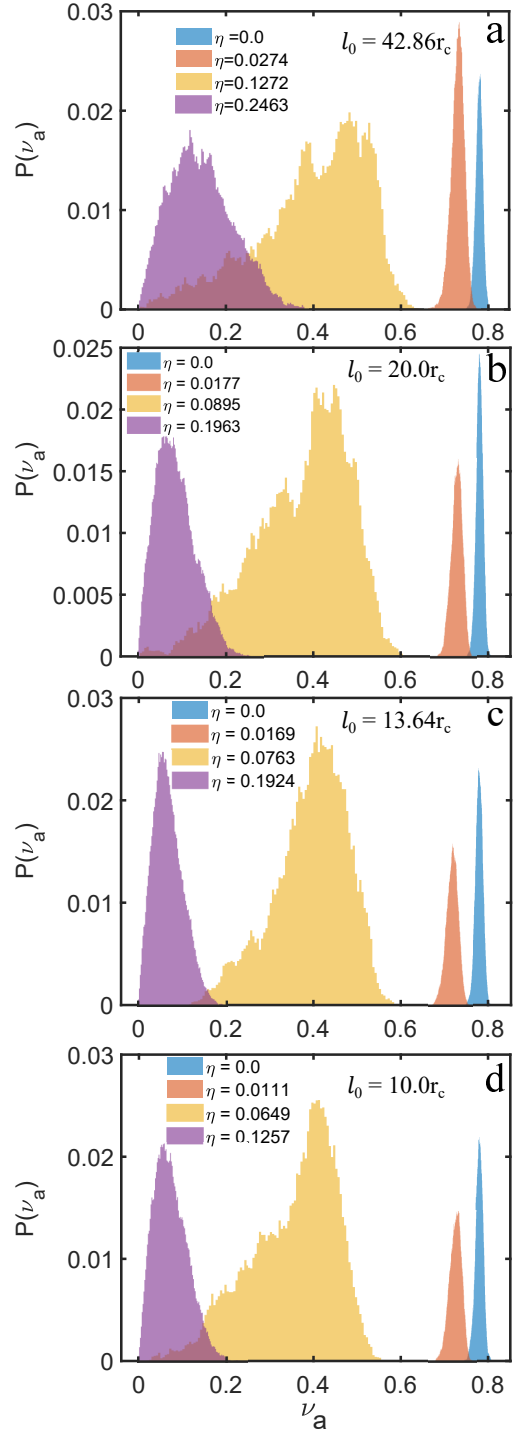}
\caption{\label{fig:Histogram} Histograms of the probability distributions of the order parameter, $P(\nu_a)$, for different packing fractions of obstacles, $\eta$. The results are presented for the arrays with $l_0 = 42.86r_c$\textbf{(a)}, $l_0 = 20.00r_c$\textbf{(b)}, $l_0 = 13.64r_c$\textbf{(c)} and $l_0 = 10.00r_c$\textbf{(d)} }   
\end{figure}

For a better insight into the nature of the transition,  we calculate the probability distribution of $\nu_a$ during long simulations of $5\times10^5$ time steps. The results are presented in Figure \ref{fig:Histogram} in the form of histograms. The histogram for the bulk and for small values of the packing fractions are narrow and peaked (see orange and blue distributions in Figures \ref{fig:Histogram} a,b,c and d).  The histograms calculated for the systems with the packing fraction of obstacles in the region of the transition are wide and often with a few distinct peaks  (See the yellow histograms in  Figure \ref{fig:Histogram} a, b, c and d).  The histograms calculated for the values of the packing fraction of obstacle after the transition to the states with small alignment of the velocities become narrower than the histograms at the transition region but wider than in the region of well aligned velocities (small values of the packing fraction). 
We also observe that the larger the lattice constant of the obstacle arrays, the wider the distribution (See the purple histograms in Figures \ref{fig:Histogram} a,b,c and d).  
\begin{figure}[hbpt]
\centering
\includegraphics[width=1.0\textwidth]{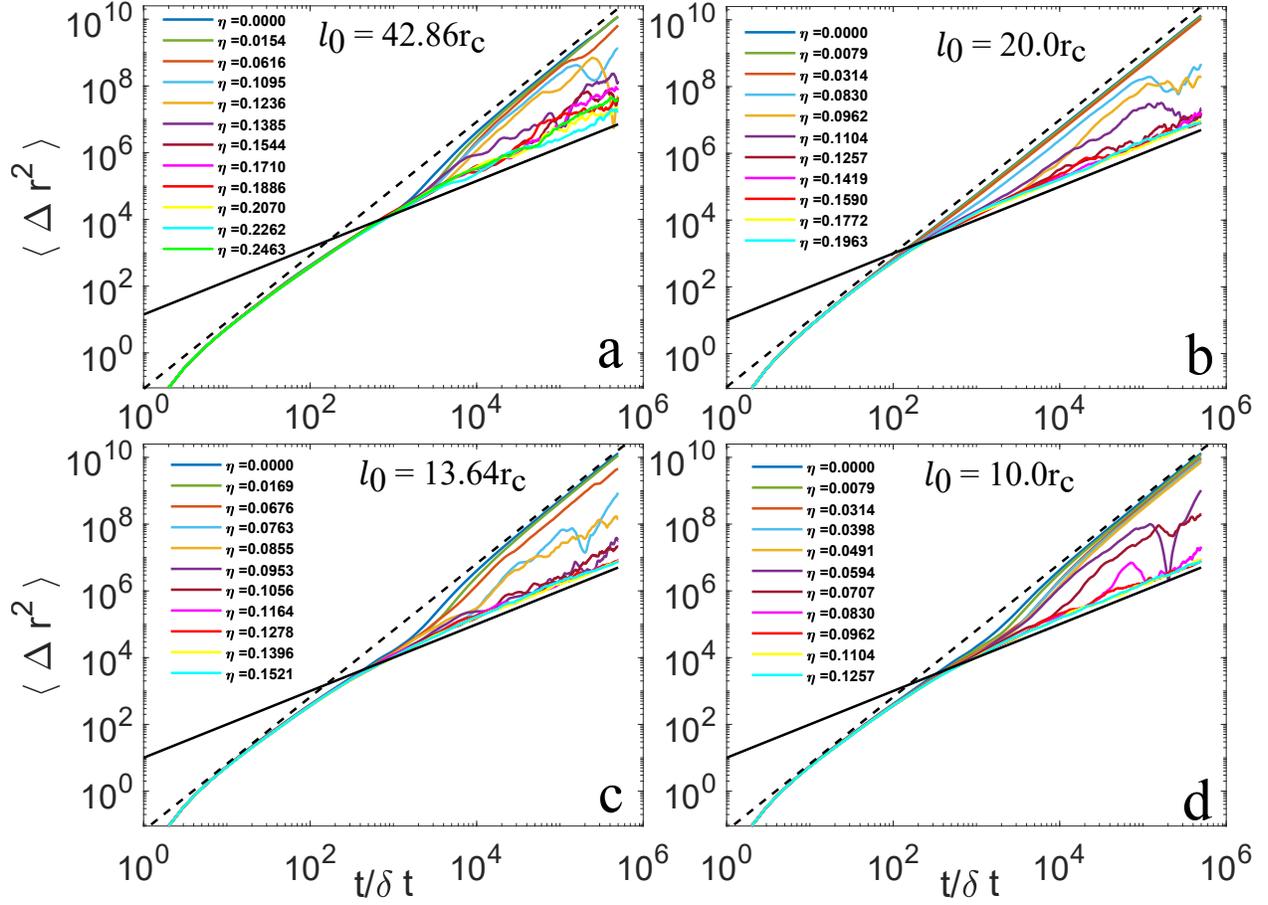}
\caption{\label{fig:MSD} Mean squared displacement, $\left\langle\Delta r(t)^2\right\rangle$ for various packing fractions of obstacles, $\eta$. The results are presented for different square lattices of obstacles with $l_0 = 42.86r_c$\textbf{(a)}, $l_0 = 20.00r_c$\textbf{(b)}, $l_0 = 13.64r_c$\textbf{(c)} and $l_0 = 10.00r_c$\textbf{(d)}. The black lines represent the limit of a pure diffusive regime (solid) and super diffusive regime (dashed)}   
\end{figure} 

Finally, to characterize the order-disorder transition from a dynamic point of view, we calculate the mean squared displacement (MSD) and the average number of collisions between point-like particles and obstacles per unit time. The MSD results are presented in Figure \ref{fig:MSD}. For all investigated systems, the behaviour at long times is super diffusive for small values of $\eta$ which correspond to the big clusters regime. However, when $\eta$ is approaching $\eta_t$, the MSD of the system starts to show a non-monotonic intermediate behaviour between super diffusive and diffusive regimes. At packing fractions about $\eta_t$, the MSD plots  indicate mixed behavior partly super diffusive, partly diffusive as shown in  Figures \ref{fig:MSD}a, b and c for $\eta = 0.1095$, $0.0830$ and $0.0763$, respectively. We have observed that in this transition region, big clusters can split into small clusters moving in different directions and  the reverse process also takes place. 
There is a correlation between the wide shape of the $\nu_a$ histograms (See Figure \ref{fig:Histogram}) and the  non-monotonic plots of the MSD. Finally, for $\eta >\eta_t$, beyond the transition region, although the MSD do not show a well developed diffusive regime, we can state that the systems approach a diffusive behaviour. This diffusive-like behaviour corresponds to regime of small clusters moving in different directions.

\begin{figure}[hbpt]
\centering
\includegraphics[width=0.5\textwidth]{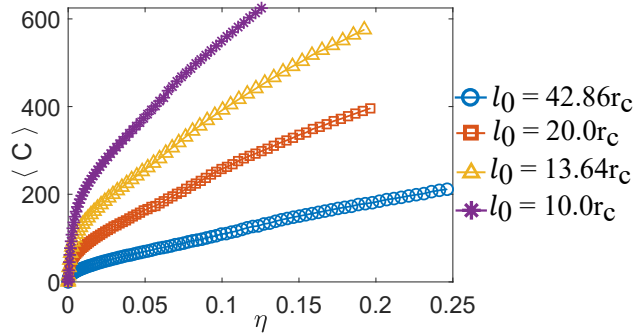}
\caption{\label{fig:Collisions} The average number of collisions with obstacles per unit time, $\langle C \rangle$, as a function of the packing fraction of obstacles, $\eta$. The error bars are comparable to the size of the markers}   
\end{figure} 

Figure \ref{fig:Collisions} shows the average number of collisions with obstacles per unit time, $\langle C\rangle$, as functions of $\eta$ for all studied systems. 
For small values of $\eta$, the slope of all curves is high, whereas the slope is smaller  for larger values of $\eta$. Thus the increase of the obstacle's radius from the values close to zero causes rapid increase of collision until some point from which the increase of collisions is substantially slower.   Near this specific point the all the curves bent significantly. 
We also observe that the number of collisions for the systems with lower density of obstacles (larger distance between obstacles) the increase of the collisions is larger when the packing fraction is increasing (which is equivalent to the increase of the size of the obstacles). 
As expected, $\langle C\rangle$ increases rapidly  as $\eta$ increases for the systems with the largest number density of obstacles ($l_0 = 10.0r_c$ and $13.64r_c$), whereas for the systems with the fewer number density of obstacles ($l_0 = 12.86r_c$ and $20.00r_c$), it increases slowly. The point where the curves bend is shifted to smaller values of $\eta$ when the system has smaller density of obstacles but it does not seem to be related with any structural or dynamical feature of the system . The location of the transition is not  clearly reflected  in the graphs of $\langle C\rangle$, since they are smooth and continuous within the whole transition region, suggesting a continuous transition.

\section{Summary and Conclusions}\label{conclusions}

We have investigated the influence of obstacles on the collective motion of large number of self propelled objects within the framework of Vicsek model. We have focused on the systems with many obstacles arranged on a square lattice. We have studied the systems with different packing fraction and number density of obstacles. We have discovered that by increasing the packing fraction of obstacles the investigated systems undergo a structural transition. The structure of the collectively moving objects changes from a few large clusters moving in a preferred direction to  many small clusters moving in different directions.  Such transition takes place for the systems with different number density of obstacles.  The change of the structure of the systems induced by the change of the packing fractions of obstacles is reflected in the changes of the order parameter,  $\nu_a$, (see Figure \ref{e:OP}) and the plots of MSD (see Figure \ref{fig:MSD}). We have determined that the transition is continuous.  We show that this transition is closely related to the dynamic features of the system. The big clusters regime corresponds to super diffusive behaviour whereas the small clusters regime is closer to diffusive behaviour. We also show that  the number density of obstacles is increased, the  transition is shifted to smaller values of $\eta$ (See Figure \ref{e:OP}a). 
The transition is also confirmed by the presence of peaks when the fluctuations of the order parameter $\Delta \nu_a$ are plotted as functions of $\eta$ (see Figure \ref{fig:OP}b). From $\Delta \nu_a$ we identify the packing fraction at which the transition occurs, $\eta_t$.
Interestingly, we have observed non-monotonic behaviour in the mean square displacement of the systems at the transition region (see Figure \ref{fig:MSD}). This behaviour results from transitions between two states, one with  a few large clusters moving in a prferred direction and the other composed of many small clusters moving in different directions.  Such behavior is reflected in the shape of the histograms of the probability of the order parameter $\nu_a$ for $\eta$ close to $\nu_t$ (see Figure \ref{fig:Histogram}). The transition described here resembles the transition in the Vicsek model induced by the increase of the noise parameter.

\begin{acknowledgement}

We would like to acknowledge the support from NCN (grant no 2018/30/Q/ST3/00434).

\end{acknowledgement}

\bibliography{sample.bib}

\end{document}